\documentstyle[12pt,aasms4]{article} 

\def\vol#1  {{{#1}{\rm,}\ }}

\def\clock{\count0=\time \divide\count0 by 60
     \count1=\count0 \multiply\count1 by -60 \advance\count1 by \time
     \number\count0:\ifnum\count1<10{0\number\count1}\else\number\count1\fi}

\begin{document}
\title{An Intrinsic Smoothing Mechanism For Gamma-Ray Burst Spectra in the Fireball Model}
\author{Renyue Cen}
\centerline{Princeton University Observatory, Princeton University, Princeton, 
NJ 08544}
\centerline{cen@astro.princeton.edu}

\begin{abstract}

It is shown that
differential Doppler shift of different patches of the blastwave front
in the fireball model
at varying angles to the line of sight
could provide an intrinsic smoothing mechanism 
for the spectra of gamma-ray bursts (GRBs)
and the associated afterglows at lower energy bands.
For the model parameters of interest,
it is illustrated that 
a monochromatic spectrum
at $\nu$ 
in blastwave comoving frame 
is smoothed and observed to have
a half-width-half-maximum (HWHM) of $\sim (0.6-2.0)\nu$.
Some other implications of this smoothing process are discussed. 
In particular, if the circumburster medium is uniform and electron and 
magnetic energies are fixed fractions of the total post-shock energy 
with time, the observed GRB or afterglow spectra cannot be steeper than
$\nu^{-\alpha}$ with $\alpha=0.75-1.25$, regardless of
the intrinsic spectra in the comoving shock frame;
i.e., a very steep electron distribution function (with $p>3$)
could produced an observed spectrum with $\alpha\sim 1$.
In addition, a generic fast-rise-slow-decay
type of GRB temporal profile is expected.

\end{abstract}


\section{Introduction}

It is remarkable that 
the simple fireball model for cosmological gamma-ray bursts
can explain the many major features of the gamma-ray bursts
and their afterglows
(Rees \& M\'ezs\'aros 1992, 1994; Paczynski \& Rhodes 1993;
Wijers, Rees, \& M\'ezs\'aros 1997;
Vietri 1997a,b;
Waxman 1997a,b;
Reichart 1997;
Katz \& Piran 1997;
Sari 1997).
It is noted, however, that the fireball model 
may produce multiple spectral components
due to the existence of both the forward and reverse shocks at least 
at some time period (e.g., M\'ezs\'aros, Rees, \& Papathanassiou 1994).
Moreover, reprocessing of the primary spectrum by, 
for example, inverse Compton scattering (e.g., Pilla \& Loeb 1998),
introduces additional features to the spectra,
even if it is  initially featureless.
Finally, resonant line features or recombination
lines due to heavy elements, for example, in dense blobs
as discussed in  M\'ezs\'aros \& Rees (1998),
may add additional features to the continuum.
In this {\it Letter} it is pointed out that
spectral smoothing due to differentially varying
Doppler shift for different patches of the fireball
at varying angles to the line of sight
provides an intrinsic, unavoidable smoothing mechanism.
While this effect is fairly well known,
a more quantitative analysis focusing on the case
of GRBs is useful. 
This {\it Letter} presents such a semi-quantitative analysis.

\section{Differential Doppler Shift Across a Fireball Front}

For the present illustration I assume that the GRB fireball
is spherical and homogeneous,
for which a single parameter, $\theta$,
the angle to the burster-observer vector,
is sufficient to 
characterize the direction of a traveling patch of 
shock heated, radiation emitting material.
It is convenient to use 
the time measured in
the rest-frame of the burster, $t$,
as the independent variable to express
other quantities.

First, one has to find the relation between time $t$ 
for a fireball patch at $\theta$ and
the time measured by
the observer on the Earth (called ``{\it O}" hereafter),
$t_{obs}$, i.e., the arrival time.
Since the {\it apparent} perpendicular traveling speed 
of a patch with $\theta$ seen by {\it O} is
\begin{equation}
\beta_{\perp} (t)={\beta (t) \sin\theta \over 1-\beta (t)\cos\theta}
\end{equation}
\noindent
in units of the speed of light,
where $\beta (t)$ is the spherical expansion speed of the fireball
in the rest-frame of the burster.
By definition, $\beta_\perp(t)$ is
\begin{equation}
\beta_{\perp}(t)={dr \over dt_{obs}}\sin\theta.
\end{equation}
\noindent
Combining equations (1, 2)
gives
\begin{equation}
{dr \over dt_{obs}} = {\beta(t)\over 1-\beta(t)\cos\theta}.
\end{equation}
\noindent
Since one also has the following relation
\begin{equation}
{dr \over dt} = \beta (t),
\end{equation}
\noindent
one finds the equation relating $t_{obs}$ to $t$:
\begin{equation}
{dt_{obs} \over dt} = 1-\beta(t)\cos\theta.
\end{equation}

Next, to have a tractable treatment, it is 
assumed that the Lorentz factor,
$\Gamma(t)\equiv 1/\sqrt{1-\beta(t)^2}$,
has the following simplified evolution:
it is constant (equal to $\Gamma_i$)
at $t\le t_{dec}$ and decays 
at $t > t_{dec}$ as
\begin{equation}
\Gamma = \Gamma_i ({t\over t_{dec}})^{-\alpha},
\end{equation}
\noindent
where $\Gamma_i$ is the initial Lorentz factor
and $t_{dec}$ (measured in the rest-frame of the burster)
characterizes
the transition time after which deceleration of the fireball
expansion becomes significant 
(hence the fireball kinetic energy can be converted into radiation)
and can be expressed approximately
as $t_{dec}=({3E\over 4\pi \Gamma_i^2 c^5 m_p n})^{1/3}$
(Blandford \& McKee 1976),
where $E$ is the initial fireball energy,
$n$ is the density of the circumburster 
medium (which is assumed to be uniform, for simplicity)
and other notations are conventional.
Note that $\alpha=3$, if the fireball cools radiatively efficiently,
and $\alpha=3/2$, if the fireball cools only adiabatically.

To make the point in a simple way it is assumed that
the total luminosity per unit frequency (in the comoving
frame) of the shock front
at a fixed time $t$ is a delta function in frequency
(i.e., a monochromatic spectrum):
\begin{equation}
L_\nu ({\nu^\prime},t) = C(t)\delta [\nu^\prime -\nu_0(t)]
\end{equation}
\noindent
in the frame comoving with the blastwave.
The characteristic frequency $\nu_0(t)$ 
at time $t$ in the comoving frame
is parameterized as
\begin{equation}
\nu_0(t) = A ({\Gamma\over\Gamma_i})^\psi,
\end{equation}
\noindent
where $A$ is a constant.
Note that, in the case of synchrotron radiation
and assuming that both the electron thermal energy density
and the magnetic energy density are fixed fractions
of the post-shock nucleon thermal energy density,
one has $\psi=3$.
$C(t)$ is expressed as
\begin{equation}
C(t) = B ({t\over t_{dec}})^\xi,
\end{equation}
\noindent
where $B$ is another constant and the $\xi$
parameterizes the temporal profile of the 
amplitude of the radiation.

At a given time {\it O}
receives radiation from {\it different parts across
the fireball surface}, emitted at varying times in the burster frame; 
i.e., radiation from regions of varying $\theta$ at varying
$t$ [see equation (5) for the relation between $t$ and $t_{obs}$]
is seen by {\it O} at the same time
(Sari 1998).
The received frequency of the radiation at $t_{obs}$
from region with $\theta$ emitted at $t$ is
\begin{equation}
\nu(t_{obs}) = \nu_0(t) D(\theta,t),
\end{equation}
\noindent
where $D(\theta,t)$ is the Doppler factor for regions with $\theta$ at time $t$:
\begin{equation}
D(\theta,t) = {1\over\Gamma(t) [1-\beta(t)\cos\theta]}.
\end{equation}
\noindent
The flux density observed by {\it O} at frequency $\nu$
at time $t_{obs}$ is 
\begin{eqnarray}
S(\nu,t_{obs}) = {1\over 8\pi d^2} \int_{-1}^1 L_\nu(\nu^\prime,t) D^3(\theta,t) d\mu 
\end{eqnarray}
\noindent
where $\nu'$($=\nu/D$) is the frequency in the blastwave frame,
$d$ is the distance of the GRB from $O$ and $\mu\equiv\cos\theta$.
Combining equations (10,11) gives
\begin{equation}
{\rm d} \mu = \left( \frac{1 - \beta \mu}{\beta} \right) 
\: \frac{{\rm d} \nu}{\nu} \; .
\end{equation}
\noindent
Inserting equations (7,13) into equation (12)
and integrating over $\nu$ give
\begin{eqnarray}
S(\nu,t_{obs}) = {C(t)\over 8\pi d^2 \nu_0(t)} D^3(\theta,t) {1-\beta(t)\cos\theta\over\beta(t)}.
\end{eqnarray}
\noindent
Note that $S(\nu,t_{obs})$ is in a parametric form;
given $t_{obs}$, one can determine the burster frame time $t$
for a given patch at $\theta$ using equation (5).
Then, one determines
$S(\nu,t_{obs})$ using 
equation (14) combined with equations (6,8,9,11), 
given $t$ and $\theta$.
Meantime, $\nu(t_{obs})$ is related to $t$ and $\theta$ by equation (10).
Thus, one can find $S(\nu,t_{obs})$ as a function of 
$\nu(t_{obs})$ at $t_{obs}$.
Let us consider a few simple but 
relevant cases to illustrate the effect.

\noindent Case 1: $\alpha=3/2$, $\psi=0$ and $\xi=0$.
This case may have some bearing on such radiation features as
atomic line features whose intrinsic 
frequencies are independent of $t$ (i.e., $\psi=0$).

\noindent Case 2: $\alpha=3$, $\psi=3$ and $\xi=1$.
This case may be related to the epoch where
radiative cooling is efficient and intensity at
the frequency in question is still rising,
which could be relevant for radiation at frequencies 
lower than the peak of the synchrotron 
radiation spectrum (due to a truncated
power-law electron distribution)
at early times of a GRB event.

\noindent Case 3: $\alpha=3$, $\psi=3$ and $\xi=-1$.
This may be related to the epoch where
radiative cooling is efficient and intensity at
the frequency in question has started to decrease,
which could be relevant for radiation at frequencies 
higher than the peak frequency of the spectrum
at early times
of a GRB event.

\noindent Case 4: $\alpha=3/2$, $\psi=3$ and $\xi=1$,
This case is similar to Case (2) with
the primary difference that 
radiative cooling is unimportant here.
This may be relevant for GRB afterglows
such as the radio afterglows when
electron cooling time is likely to be significantly longer
than the dynamic time of the expanding fireball
at frequencies lower than the peak frequency of the spectrum.

\noindent Case 5: $\alpha=3/2$, $\psi=3$ and $\xi=-1$,
This case is similar to Case 4 
but for frequencies higher than the peak frequency of the spectrum.

While it is convenient to express various quantities 
using $t$ as the independent time variable,
one needs to express the final observables using $t_{obs}$,
which is related to $t$ (for $\theta=0$; see equation 5) as
\begin{equation}
t_{obs} = {t_{dec}\over 2\Gamma_i^2}\left[1+{1\over 2\alpha+1} \left({t\over t_{dec}}\right)^{2\alpha +1}\right]
\end{equation}
\noindent
for the simplified solution of $\Gamma(t)$ given by equation (6).
Figure (1a) shows the flux density as a function of
frequency at $t_{obs}=t_{dec}/\Gamma_i^2$ 
for the five cases.
The frequency is normalized such that unity
corresponds to the radiation from regions with $\theta=0$
and the flux density is normalized to be unity at unity frequency.
The sharp turns to the left for cases (ii,iii,iv,v)
correspond to the sharp turn
of the evolution of the Lorentz factor at $t=t_{dec}$.
Figure (1b) shows the flux density as a function of
frequency at $t_{obs}=5000t_{dec}/\Gamma_i^2$ 
for the five cases.
Also shown in both panels are two straight lines in the upper
right corner indicating the spectral slope of $-0.75$ and $1.25$,
respectively, which bracket the range of the spectral slope
for various cases shown.
Note that sharp turns as seen in (1a) are not visible simply because
they appear at much lower intensity level than the displayed range
in the figure.
Note that
$t_{dec}=16\hskip -0.1cm\left({E\over 10^{52}{\rm erg}}\right)^{1/3}\hskip -0.2cm \left({\Gamma_i\over 300}\right)^{-2/3}\hskip -0.2cm \left({n\over 1 {\rm cm}^{-3}}\right)^{-1/3}\hskip -0.1cm{\rm days}$.
Therefore, for typical values of $E$,
$\Gamma_i$ and $n$,
$t_{obs}$ is of order a second and a day, respectively,
after the fireball
explosion for the two cases shown in (1a) and (1b).
These two cases may respectively be relevant for bursts
in gamma-ray and afterglows at lower energy bands. 
It should be noted that, although the external shock model
is used to illustrate the smoothing magnitude,
the results should be applicable to the internal shock model as well.

Recall that in all cases a delta function spectrum 
at $\nu$ is assumed
in the comoving frame at a given time.
One sees
that this delta function spectrum is smoothed out to appear
as a broad spectrum with 
a half-width-half-maximum (HWHM) of $\sim (0.6-2.0)\nu$
for all the cases considered, except for Case 1 where 
HWHM is $\sim 0.1\nu$.
An immediate implication from Figure 1 (see the dot-long-dashed curves
on the upper right corner in the two panels)
is that the observed spectra
of GRBs or their afterglows should roughly have $\nu^{-1}$,
if the electron distribution function power index $p$ is equal to 
greater than $3$.
In other words,
{\it the observed spectra cannot be steeper than $\nu^{-1}$ 
regardless the value of $p$}.
This slope 

It is of interest to understand where the different
radiation that correspond to different frequencies in Figure 1
comes from.
Figure 2 shows the frequency seen by
{\it O} at a given time as a function of
the emission shell radius from the burster, $r$, 
expressed in units of $c t_{dec}$.
It is  seen that, for realistic cases (ii,iii,iv,v) that
correspond to the continuum radiation of the GRBs and afterglows,
higher frequency radiation comes from earlier
time $t$ (in the burster frame) with larger angle $\theta$
up to $t_{dec}$ after which there is a downturn
to lower frequency at still earlier times
due to assumed constancy of $\Gamma$ thus constancy
of $\nu_0(t)$.
For case (i) lower frequency radiation comes from 
earlier time $t$ with larger $\theta$.
In both panels (a,b) also shown
along the solid curves using solid dots
are the corresponding $\theta$ values
in degrees.
Dotted and dashed curves are also punctuated
by open circles and open squares,
corresponding to the same $\theta$ values.

\section{Discussion}

It is shown that differentially varying Doppler boost of different
patches of the fireball front
provides an intrinsic, unavoidable smoothing mechanism for the 
spectra of gamma-ray bursts and their afterglows.
The detailed smoothing patterns are complicated,
depending upon various factors such as
the evolution of the Lorentz factor and 
the evolution of the intrinsic (i.e., comoving frame)
radiation spectrum.
Nonetheless, for plausible ranges of model parameters of interest,
a comoving frame delta function spectrum at $\nu$ is smoothed to have
a HWHM of $\sim (0.6-2.0)\nu$, assuming that
the time evolution of the characteristic frequency $\nu(t)$
is proportional to some positive power of
the shock front Lorentz factor,
$\Gamma^\psi$ (where $\psi>0$; see equation 8). 
This type of smoothing may be applicable to continuous
spectra such as from synchrotron mechanism.

In the case $\psi=0$ (appropriate
for atomic line features which
are independent of the blastwave dynamics),
the spectral smoothing is smaller,
with a HWHM of $\sim 0.1\nu$.
Thus, a sharp linelike emission feature (in the comoving frame)
would be smoothed out to have
an equivalent width of about $0.1\nu$.
Furthermore, the spectral profile of such a feature
will be asymmetrical with 
a sharp cutoff at the high end (see the
two solid curves in Figure 1).

Two interesting and natural consequences arise due to the differential Doppler
smoothing. 
First, the observed spectra
of GRBs or their afterglows cannot be steeper than $\nu^{-\alpha}$
with $\alpha=0.75-1.25$ (Figure 1),
even though the intrinsic spectra in the comoving shock frame
may be much steeper, if the circumburster medium is uniform and electron and 
magnetic energies are fixed fractions of the total post-shock energy with time.
Second, a generic
fast-rise-slow-decay type temporal profile
of the GRB bursts is expected,
not necessarily reflecting the intrinsic temporal
profiles of the bursts in the comoving frame.
This can be easily seen by considering 
the case where the intrinsic (blastwave frame)
spectrum is a bivariate delta function in both time and frequency.
In this case the fast rise occurs when
the radiation from the region around $\theta=0$ enters the 
observer's finite band.
Subsequently, radiation from 
regions with gradually increasing $\theta$
is received at decreasing amplitudes 
in the same observer's finite band 
with a decaying time scale of $\sim t_{dec}/2\Gamma^2$
(in observer's frame).

\acknowledgments
The work is supported in part
by grants AST9318185 and ASC9740300.
I thank Bohdan Paczy\'nski for discussion.

\newpage
\figcaption[FLENAME]{
Panels (a,b) shows the flux density as a function of
frequency for the five cases (see text).
at $t_{obs}=t_{dec}/\Gamma_i^2$ 
and $t_{obs}=5000t_{dec}/\Gamma_i^2$, respectively. 
A fiducial value of $\Gamma_i=300$ is used.
In both panels there are two straight lines in the upper
right corner showing the spectral slope of $-0.75$ and $1.25$,
respectively, which approximately bracket the range of the spectral slope
for various cases shown.
\label{fig1}}

\figcaption[FLENAME]{
shows the frequency seen by
{\it O} at a given time as a function of
the emission shell radius from the burster $r$, expressed
in units of $c t_{dec}$
for the five cases, 
with panel (a) at $t_{obs}=t_{dec}/\Gamma_i^2$ 
and panel (b) at $t_{obs}=5000t_{dec}/\Gamma_i^2$.
A fiducial value of $\Gamma_i=300$ is used.
In both panels also shown
along the solid curves using solid dots
are the corresponding $\theta$ values in degrees.
Dotted and dashed curves are also punctuated
by open circles and open squares,
corresponding to the same $\theta$ values.
\label{fig2}}

\clearpage

\begin{thebibliography}{DUM}
\bibitem[Blandford \& McKee 1976]{bm76} Blandford, R.D., \& McKee, C.F. 1976, Phys. Fluids, 19, 1130
\bibitem[Katz \& Piran 1997]{kp97} Katz, J., \& Piran, T. 1997,\apj, 490, 772
\bibitem[M\'esz\'aros \& Rees 1998]{mr98} M\'esz\'aros, P., \& Rees, M.J. 1998, ApJ, 502, L105
\bibitem[M\'esz\'aros, Rees, \& Papathanassiou 1994]{mrp94} M\'esz\'aros, P., Rees, M.J., \& Papathanassiou, H. 1994, ApJ, 432, 181
\bibitem[Paczy\'nski \& Rhodes 1993]{pr93} Paczy\'nski, B., \& Rhodes, J. 1993, \apj, 418, L5
\bibitem[Pilla \& Loeb 1998]{pl98} Pilla, R.P., \& Loeb, A. 1998, ApJ, 495, 597
\bibitem[Rees \& M\'esz\'aros 1992]{rm92} Rees, M.J., \& M\'esz\'aros, P. 1992, \mnras, 258, 41 
\bibitem[Rees \& M\'esz\'aros 1994]{rm94} Rees, M.J., \& M\'esz\'aros, P. 1994, \apj, 430, L93
\bibitem[Reichart 1997]{r97} Reichart, D.E. 1997, \apj, 485, L57
\bibitem[Sari 1997]{sa97} Sari, R. 1997, \apj, 489, L37
\bibitem[Sari 1998]{sa98} Sari, R. 1998, \apj, 494, L49
\bibitem[Vietri 1997a]{vi97a} Vietri, M. 1997a, \apj, 478, L9
\bibitem[Vietri 1997b]{vi97b} Vietri, M. 1997b, \apj, 488, L105
\bibitem[Waxman 1997a]{w97a} Waxman, E. 1997a, \apj, 485, L5 
\bibitem[Waxman 1997b]{w97b} Waxman, E. 1997b, \apj, 489, L33 
\bibitem[Wijers, Rees, \& M\'ezs\'aros 1997]{wrm1997}Wijers, A.M.J., Rees, M., \& M\'ezs\'aros, P. 1997, \mnras, 288, L51
\end{thebibliography}
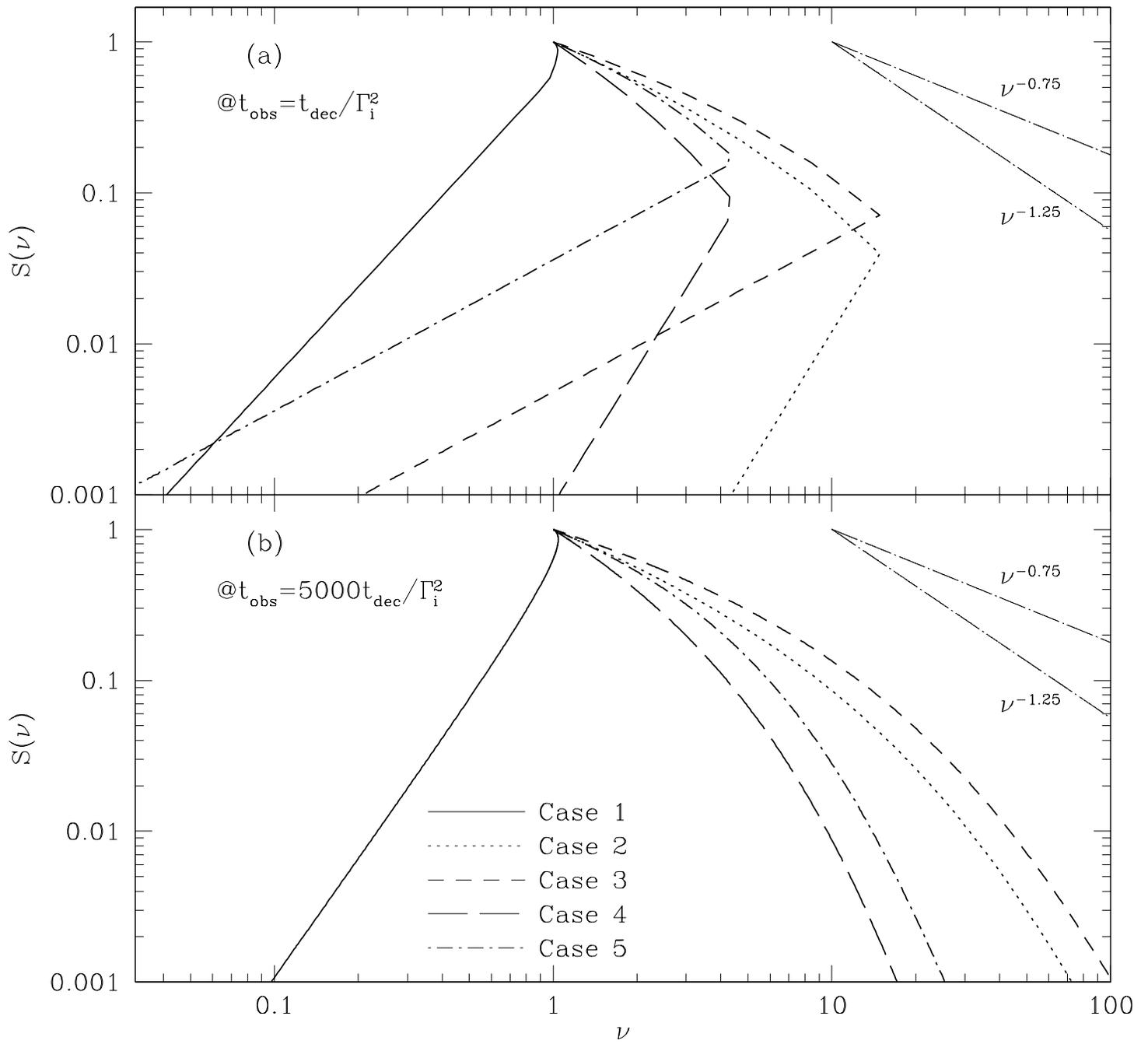
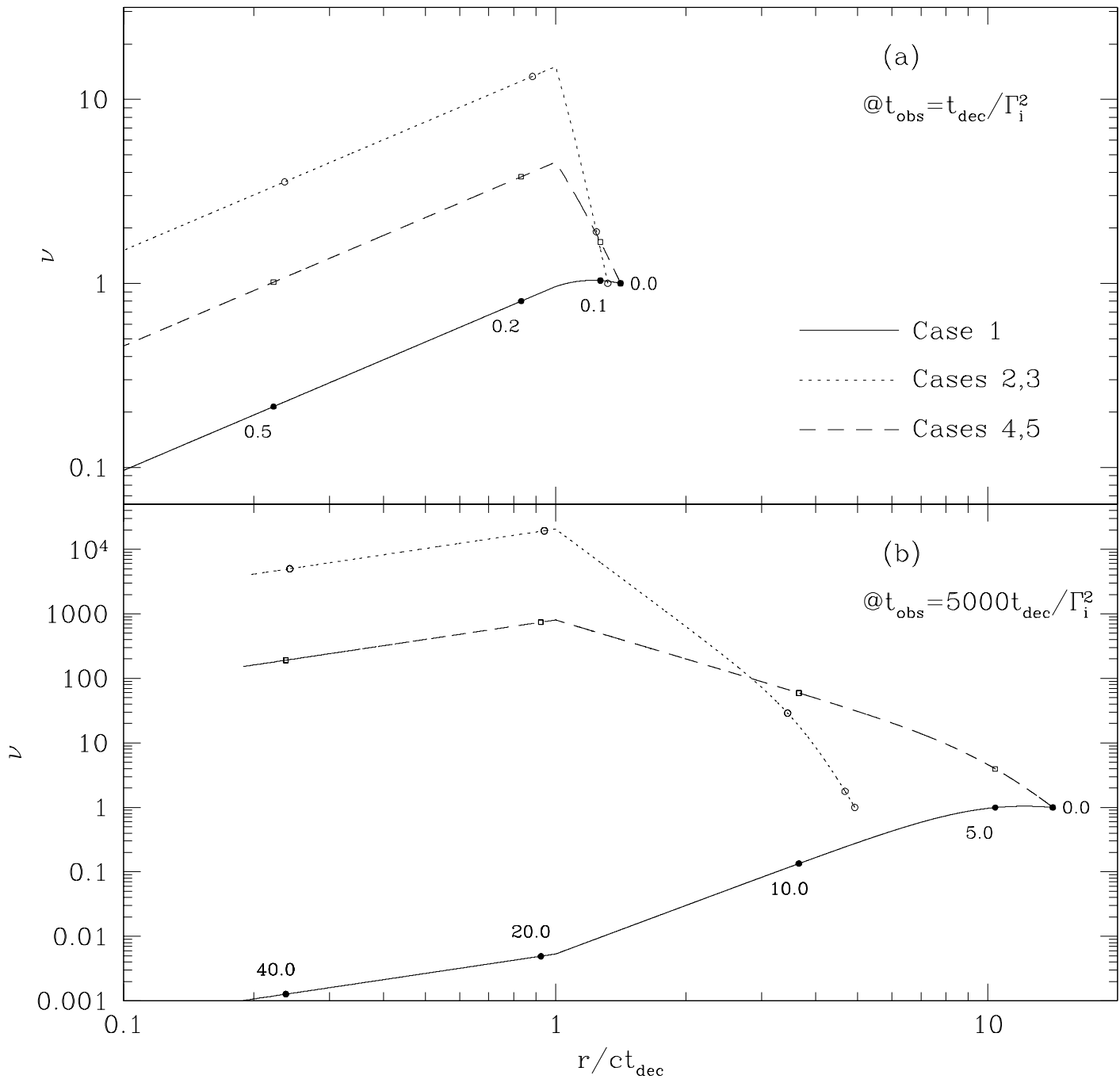
\end{document}